\documentclass[aps,prl,twocolumn,groupedaddress]{revtex4}

\usepackage{graphicx}
\usepackage{dcolumn}
\usepackage{bm}

\newcommand{\be}{\begin{equation}}
\newcommand{\ee}{\end{equation}}
\newcommand{\ba}{\begin{eqnarray}}
\newcommand{\ea}{\end{eqnarray}}
\newcommand{\baa}{\begin{eqnarray*}}
\newcommand{\eaa}{\end{eqnarray*}}
\newcommand{\lab}[1]{\label{#1}}

\newcommand{\dis}{\displaystyle}
\newcommand{\biq}{\mbox{\boldmath $q$}}

\newcommand{\biL}{\mbox{\boldmath $\Lambda$}}
\newcommand{\bil}{\mbox{\boldmath $\lambda$}}

\newcommand{\bs}[1]{\mbox{\boldmath $#1$}}
\begin{document}


\title{Multicanonical algorithm, simulated tempering, replica-exchange method, and all that}

\author{Ayori Mitsutake}
\email{ayori@mail.rk.phys.keio.ac.jp}
\affiliation{ 
Department of Physics, Keio University, Yokohama, Kanagawa 223-8522, Japan}
\author{Yuko Okamoto}%
 \email{okamoto@phys.nagoya-u.ac.jp}
\affiliation{%
Department of Physics, Nagoya University, Nagoya, Aichi 464-8602, Japan}


\begin{abstract}
We discuss multi-dimensional generalizations of multicanonical algorithm, 
simulated tempering, and replica-exchange method. 
We generalize the original potential energy function $E_0$ by
adding any physical quantity $V$ of interest as a new
energy term with a coupling constant $\lambda$.
We then perform a multi-dimensional multicanonical simulation
where a random walk in $E_0$ and $V$ space is realized.
We can alternately perform a multi-dimensional simulated tempering
simulation where a random walk in temperature $T$ and 
parameter $\lambda$ is
realized.
The results of the multi-dimensional replica-exchange simulations
can be used to determine the weight factors for 
these multi-dimensional multicanonical
and simulated tempering simulations.
\end{abstract}

\maketitle

Monte Carlo (MC) and molecular dynamics (MD) simulations of 
frustrated systems such as spinglass and
biomolecular systems are very difficult because their free energy landscapes 
are rugged and long equilibration time is necessary. 
In order to 
overcome this difficulty, {\it generalized-ensemble algorithms} have often
been employed (for reviews, see, for instance, Refs.~\cite{RevHO1,Be02,RevO2,RevSMO}).
Generalized-ensemble algorithms are based on artificial, non-Boltzmann weight
factors so that random walks in potential energy space and other variable
space may be realized. Once an optimal weight factor is found, one makes
a single long production run. From the results of this production run,
one can reconstuct canonical, realistic  ensembles for a wide range of
temperature and other parameter values by the single-histogram \cite{FS1}
or multiple-histogram \cite{FS2,WHAM} reweighting techniques.
{\it Multicanonical algorithm} (MUCA)\cite{MUCA2,MUCA3}, 
{\it simulated tempering} (ST)\cite{ST1,ST2},
and {\it replica-exchange method} (REM)\cite{RE1} are three of the most widely
used generalized-ensemble algorithms. (ST is also referred to as the 
{\it method of expanded ensemble} \cite{ST1} and REM is also referred to as \
{\it parallel tempering} \cite{STrev}.)
In this article, we present general formulations for multi-dimensional
extensions of these three methods, where we generalize the original potential energy function by adding any physical quantity of interest as a new energy term so that a random walk not only in the original potential energy space but also in the additional energy space is realized. 
 
Let us consider the following generalized potential
energy function of a system in state $x$:
\begin{equation}
E_{\bil} (x) = E_0 (x) + \sum_{\ell = 1}^L
\lambda^{(\ell)} V_{\ell} (x)~.
\label{eqn1}
\end{equation}
Here, there are $L+1$ energy terms, $E_0(x)$ and
$V_{\ell}(x)$ ($\ell =1, \cdots, L$), and
$\lambda^{(\ell)}$ are the
corresponding coupling constants for 
$V_{\ell}(x)$ 
(we collectively write $\bil = (\lambda^{(1)}, \cdots, \lambda^{(L)})$).
The partition function of the system at fixed temperature $T$ and $\bil$ 
is then given by
\begin{equation}
\begin{array}{ll}
\lefteqn{Z(T,\bil)=\int dx \exp(-\beta E_{\bil}(x))}\\
 &=\dis \int dE_0 dV_1 \cdots dV_L ~ n(E_0,V_1,\cdots,V_L) 
\exp \left(-\beta E_{\bil} \right)~,
\end{array}
\label{eqn2}
\end{equation}
where $\beta = 1/ k_{\rm B} T$, $k_{\rm B}$ is the Boltzmann constant, and 
$n(E_0,V_1,\cdots,V_L)$ is the multi-dimensional density of states. 
Here, the integral is replaced by a summation when $x$ is discrete. 

The expression in Eq.~(\ref{eqn1}) is 
often used in simulations. 
For instance, 
in simulations of spin systems, 
$E_0(x)$ and $V_{1}(x)$ (here, $L = 1$ 
and $x=\{ S_1, S_2, \cdots \}$ stand for spins) 
can be respectively considered as the zero-field term 
and the magnetization term coupled with
the external field $\lambda^{(1)}$. 
(For Ising model, $E_0 = -J \sum_{<i,j>} S_i S_j$, 
$V_1 = - \sum_i S_i$, and $\lambda^{(1)}=h$, i.e., external magnetic field.) 
In umbrella sampling \cite{US} in molecular simulations, $E_0(x)$ and
$V_{\ell}(x)$ can be taken as the original potential
energy and the ``biasing'' umbrella potential energy, respectively, with the
coupling parameter $\lambda^{(\ell)}$ (here, $x= \{{\biq}_1, \cdots, {\biq}_N\}$ where ${\biq}_i$ are the coordinate
vectors of the $i$-th particle and $N$ is the total number of particles).  
For the molecular simulations in the isobaric-isothermal ensemble,
$E_0(x)$ and $V_1(x)$ (here, $L = 1$) can be respectively considered as the
potential energy $U$ and the volume ${\cal V}$ coupled with the pressure 
${\cal P}$. (Namely, we have
$x= \{{\biq}_1, \cdots, {\biq}_N, {\cal V}\}$, 
$E_0=U$, $V_1={\cal V}$, and
$\lambda^{(1)}={\cal P}$, i.e., $E_{\bil}$ is the enthalpy without the kinetic energy contributions.)
For simulations in the grand canonical ensemble with $N$ particles,
we have $x= \{{\biq}_1, \cdots, {\biq}_N, N \}$, and 
$E_0(x)$ and $V_1(x)$ (here, $L = 1$) can be respectively considered as the
potential energy $U$ and the total number of particles $N$ coupled with the 
chemical potential $\mu$. (Namely, we have
$E_0=U$, $V_1=N$, and
$\lambda^{(1)}=-\mu$.)
We remark that generalized-ensemble algorithms in various ensembles are 
also discussed in Refs.~\cite{OO,Esco}.
Moreover, we can introduce any physical quantity of interest (or its function) as the additional potential energy term $V_{\ell}$. For instance, $V_{\ell}$ can be an overlap with a reference configuration in spinglass systems, an end-to-end distance and a radius of gyration in molecular systems, etc. 
In such a case, we have to carefully choose the range of $\lambda^{(\ell)}$ 
values so that the new energy term $\lambda^{(\ell)} V_{\ell}$ 
will have roughly the same order of magnitude as the original energy term $E_0$. 
We want to perform a simulation where a random walk not only in the $E_0$ space but also in the $V_{\ell}$ space is realized. As shown below, this can be done by performing a multi-dimensional MUCA or ST simulation. 

We first describe the multi-dimensional MUCA simulation which realizes 
a random walk in the $( L+1 )$-dimensional space of
$E_0(x)$ and $V_{\ell}(x)$ ($\ell~=~1, \cdots, L$).
In the multi-dimensional MUCA ensemble, 
each state is weighted by the MUCA weight factor 
$W_{\rm mu}(E_0,V_1,\cdots, V_L)$ so that a uniform energy distribution of 
$E_0$, $V_1, \cdots$, and $V_L$ may be obtained:
\begin{equation}
\begin{array}{ll}
\lefteqn{P_{\rm mu}(E_0,V_1,\cdots,V_L) }\\
&\propto n(E_0,V_1,\cdots,V_L) W_{\rm mu}(E_0,V_1,\cdots,V_L) \equiv {\rm const}~,
\end{array}
\label{eqn3}
\end{equation}
where $n(E_0,V_1,\cdots,V_L)$ is the multi-dimensional density of states.
From this equation, we obtain
\begin{equation}
\begin{array}{ll}
\lefteqn{W_{\rm mu}(E_0,V_1,\cdots,V_L) \propto \frac{1}{n(E_0,V_1,\cdots,V_L)} }\\
\equiv & \exp \left(-\beta_a E_{\rm mu}(E_0,V_1,\cdots,V_L) \right)~,
\end{array}
\label{eqn4}
\end{equation}
where in the second line we have introduced an arbitrary reference temperature, 
$T_a = 1/k_{\rm B} \beta_a$, and wrote the weight factor 
in the Boltzmann-like form. 
Here, the ``{\it multicanonical potential energy}'' is defined by
\begin{equation}
E_{\rm mu}(E_0,V_1,\cdots,V_L)  \equiv k_{\rm B} T_a \ln n(E_0,V_1,\cdots, V_L)~.
\label{Eqn5}
\end{equation}
The multi-dimensional MUCA MC simulation can be performed with the
following Metropolis transition probability from state $x$ with 
energy $E_{\bil}=E_0 + \sum_{\ell=1}^{L} \lambda^{(\ell)} V_{\ell}$ to state $x^{\prime}$ with
energy ${E_{\bil}}^{\prime}={E_0}^{\prime} + \sum_{\ell=1}^{L} \lambda^{(\ell)} {V_{\ell}}^{\prime}$ :
\begin{equation}
\begin{array}{lll}
 w(x \rightarrow x^{\prime})
&=& {\rm min} \left(1,\dis \frac{W_{\rm mu}({E_0}^{\prime},{V_{1}}^{\prime},\cdots,{V_{L}}^{\prime})}
{W_{\rm mu}(E_0,V_1,\cdots,V_{L})}\right) \\
&=& {\rm min} \left(1,\dis \frac{n(E_0,V_1,\cdots,V_L)}{n({E_0}^{\prime},{V_{1}}^{\prime},\cdots,{V_{L}}^{\prime})} \right)~.
\end{array}
\label{eqn6}
\end{equation}
An MD algorithm in the multi-dimensional MUCA ensemble
also naturally follows from Eq.~(\ref{eqn4}), in which a
regular constant temperature MD simulation
(with $T=T_a$) is performed by replacing the total potential energy 
$E_{\bil}$ by the multicanonical potential energy $E_{\rm mu}$ in the Newton's equations for the $k$-th particle  ($k=1,\cdots,N$)
(see Refs. \cite{HOE96,NNK} for one-dimensional version):  
\begin{equation}
\dot{\bs{p}}_k ~=~ - \frac{\partial E_{\rm mu}( E_0,V_1,\cdots,V_L)}{\partial \bs{q}_k}~.
\label{eqn7}
\end{equation}

Secondly, we consider a multi-dimensional ST simulation which realizes a random walk both in temperature $T$ and in parameters $\bil$. 
The parameter set $\biL=(T,\bil) \equiv (T, \lambda^{(1)}, \cdots, \lambda^{(L)})$ 
become dynamical variables and both the configuration and the parameter set 
are updated during the simulation with a weight factor:
\begin{equation}
W_{\rm ST} (\biL)
\equiv \exp \left(-\beta E_{\bil} + f(\biL) \right)~,
\label{eqn8}
\end{equation}
where the function $f(\biL)=f(T,\bil)$ is chosen so that 
the probability distribution of $\biL$ is flat:
\begin{equation}
\begin{array}{ll}
&\lefteqn{P_{\rm ST}(\biL)\propto} \\
&\dis \int dE_0 dV_1 \cdots dV_L~ n(E_0,V_1,\cdots,V_L)~ 
\exp \left(-\beta E_{\bil} + f(\biL) \right) \\
&\lefteqn {\equiv {\rm const}~.} 
\end{array}
\lab{eqn9}
\end{equation}
This means that $f(\biL)$ is the dimensionless (``Helmholtz'') free energy: 
\begin{equation}
\begin{array}{ll}
\lefteqn{ \exp \left( -f(\biL) \right) } \\
&\propto \dis \int dE_0 dV_1 \cdots dV_L ~ n(E_0,V_1,\cdots,V_L)~ \exp (-\beta E_{\bil})~. 
\end{array}
\lab{eqn10}
\end{equation}

In the numerical work we discretize the parameter
set $\biL$ in $M (= M_0 \times M_1  \times \cdots \times M_L)$ different values: 
$\biL_{m} \equiv (T_{m_0},\bil_{m}) \equiv (T_{m_0},\lambda^{(1)}_{m_1},\cdots,\lambda^{(L)}_{m_L})$, where $m_0=1, \cdots, M_0, m_{\ell}=1, \cdots, M_{\ell}$ ($\ell = 1, \cdots, L$).  
Without loss of generality we can order the parameters
so that $T_1 < T_2 < \cdots < T_{M_0}$ and
$\lambda^{(\ell)}_{1} < \lambda^{(\ell)}_{2} < \cdots < \lambda^{(\ell)}_{M_{\ell}}~$ (for each $\ell = 1,\cdots, L$).  
The free energy $f$ is now written as $f_{m_0,m_1,\cdots,m_L}=f(T_{m_0},\lambda^{(1)}_{m_1},\cdots, \lambda^{(L)}_{m_L}$).

Once the initial configuration and
the initial parameter set are chosen,
the multi-dimensional ST is realized by alternately 
performing the following two steps:
\begin{enumerate}
\item A canonical MC or MD simulation at the fixed parameter
set $\biL_m = (T_{m_0},\bil_{m}) = (T_{m_0},\lambda^{(1)}_{m_1},\cdots, \lambda^{(L)}_{m_L})$ is carried out for a certain steps with the weight factor $\exp(-\beta_{m_0} E_{\bil})$.
\item 
We update the parameter set $\biL_m$ to a new parameter set $\biL_{m \pm 1}$ in which one of the parameters in $\biL_{m}$ is changed to a neighboring value with the configuration and the other parameters fixed. 
The transition probability of
this parameter-updating
process is given by the following Metropolis criterion:
\begin{equation}
\begin{array}{ll}
w(\biL_{m} \rightarrow \biL_{m \pm 1})
&= {\rm min}\left(1,\dis \frac{W_{\rm ST}(\biL_{m \pm 1})}
{W_{\rm ST}(\biL_{m})} \right) \\
&= {\rm min}\left(1,\exp \left( - \Delta \right)\right)~.
\label{eqn11}
\end{array}
\end{equation}
Here, there are two possibilities for $\biL_{m \pm 1}$, and we have 
$\biL_{m \pm 1} = (T_{m_0 \pm 1}, \cdots, \lambda^{(\ell)}_{m_{\ell}}, \cdots)$ 
with 
\begin{equation}
\Delta = \left(\beta_{m_0 \pm 1} - \beta_{m_0} \right) E_{\bil_m}
- \left(f_{m_0 \pm 1,m_1,\cdots,m_L} - f_{m_0,m_1,\cdots,m_L} \right),
\label{eqn12}
\end{equation}
for $T$-update, and 
$\biL_{m \pm 1} = (T_{m_0}, \cdots, \lambda^{(\ell)}_{m_{\ell} \pm 1}, \cdots)$ with 
\begin{equation}
\begin{array}{rl}
\Delta = \beta_{m_0} (\lambda^{(\ell)}_{m_{\ell} \pm 1} - \lambda^{(\ell)}_{m_{\ell}}) V_{\ell}
- \left(f_{m_0,\cdots,m_{\ell} \pm 1,\cdots} - f_{m_0,\cdots,m_{\ell},\cdots} \right),
\end{array}
\label{eqn13}
\end{equation}
for $\lambda^{(\ell)}$-update (for one of $\ell=1,\cdots,L$).
\end{enumerate}
 
We remark that the random walk in $E_0$ and in $V_{\ell}$ for the MUCA simulation corresponds to that in $\beta$ and in $\beta \lambda^{(\ell)}$ for the ST simulation:
\begin{equation}
\left\{
\begin{array}{ll}
E_0 & \longleftrightarrow \beta~, \cr
V_{\ell} & \longleftrightarrow \beta \lambda^{(\ell)}~,~~(\ell=1,\cdots,L)~.
\end{array}
\right.
\end{equation}
They are in conjugate relation. 

We can perform the multi-dimensional MUCA and ST simulations when we have optimal weight factors. However, we do not know these MUCA and ST weight factors {\it a priori} and need to estimate them by short preliminary simulations.
For one-dimensional version, three methods are well-known to obtain the weight factors: The first one is to use recursion formulas \cite{Be02}, 
the second one is to use Wang-Landau methods \cite{Landau1}, and the third one is to use a short REM simulation and the multiple-histogram reweighting techniques \cite{SO3,MSO03,MSO03b,MO4,STREM}.
Here, we generalize this third method to multi-dimensional versions (see also Refs. \cite{RevO2,RevSMO}).  

We use the {\it multi-dimensional replica-exchange method} (MREM) \cite{SKO} to determine the multi-dimensional MUCA and ST weight factors.
The system for MREM consists of $M$ non-interacting replicas of the original system in the
``canonical ensemble'' with $M (=M_0 \times M_1 \times \cdots \times M_L)$ different parameter 
sets $\biL_m$ ($m=1,\cdots,M$).
Because the replicas are non-interacting, the weight factor 
is given by the product of Boltzmann factor for each replica:
\begin{equation}
W_{\rm MREM}
\equiv \dis{\prod_{m=1}^M \exp \left(- \beta_{m_0} 
E_{\bil_m}
 \right)}~.
\label{eqn16}
\end{equation}

REM closely follows the ST procedures described above.
In step 1, a ``canonical'' MC or MD simulation at the fixed
parameter set is carried out for each replica 
simultaneously and independently for a certain MC or MD steps.
In step 2, 
we exchange a pair of replicas $i$ and $j$ which are at the
parameter sets $\biL_m$ and $\biL_{m+1}$, respectively.
The transition probability for this replica exchange process is given by 
\begin{equation}
w(\biL_m \leftrightarrow \biL_{m+1})
={\rm min}\left(1,\exp(-\Delta)\right),
\label{eqn18}
\end{equation}
where we have
\begin{equation}
\Delta = \left(\beta_{m_0} - \beta_{m_0+1} \right) 
\left(E_{\bil_m} \left(q^{[j]}\right) - E_{\bil_m} \left(q^{[i]}\right) \right)~,
\label{eqn31}
\end{equation}
for $T$-exchange, and 
\begin{equation}
\Delta = \beta_{m_0} \left(\lambda_{m_{\ell}}^{(\ell)}
- \lambda_{m_{\ell}+1}^{(\ell)} \right)
\left(V_{\ell}\left(q^{[j]}\right) - V_{\ell}\left(q^{[i]}\right) \right)~,
\label{eqn32}
\end{equation}
for $\lambda^{(\ell)}$-exchange (for one of $\ell=1,\cdots,L$).
Here, $q^{[i]}$ and $q^{[j]}$ stand for configuration variables
for replicas $i$ and $j$, respectively, before the replica exchange.
Usually, $M_0/2$ or $M_{\ell}/2$ pairs of replicas corresponding to neighboring $T$ or $\lambda^{(\ell)}$ are 
simultaneously exchanged, and the pairing is alternated between the two possible choices, 
i.e., $(T_1,T_2),(T_3,T_4),\cdots$ and $(T_2,T_3),(T_4,T_5),\cdots$ or 
$(\lambda_1^{(\ell)},\lambda_2^{(\ell)}),(\lambda_3^{(\ell)},\lambda_4^{(\ell)}),\cdots$ and 
$(\lambda_2^{(\ell)},\lambda_3^{(\ell)}),(\lambda_4^{(\ell)},\lambda_5^{(\ell)}),\cdots$, respectively. 

To obtain the canonical distributions,
the multiple-histogram reweighting techniques \cite{FS2,WHAM}
are particularly useful.
Suppose we have made a single run of the MREM simulation with $M (= M_0 \times M_1  \times \cdots \times M_L)$ replicas that correspond
to $M$ different parameter sets
$\biL_m$ ($m=1, \cdots, M$).
Let $N_{m_0,m_1,\cdots,m_L}(E_0,V_1,\cdots,V_L)$ and $n_{m_0,m_1,\cdots,m_L}$
be respectively 
the ($L+1$)-dimensional potential-energy histogram and the total number of
samples obtained for the $m$-th parameter set 
$\biL_m=(T_{m_0},\lambda^{(1)}_{m_1},\cdots,\lambda^{(L)}_{m_L})$. 
The multiple-histogram reweighting equations are then given by \cite{FS2,WHAM}
\begin{equation}
\begin{array}{ll}
\lefteqn{n(E_0,V_1,\cdots,V_L) }  \\
&= \frac{\dis{\sum_{m_0,m_1,\cdots,m_L} 
N_{m_0,m_1,\cdots,m_L}(E_0,V_1,\cdots,V_L)}} 
{\dis{\sum_{m_0,m_1,\cdots,m_L} 
n_{m_0,m_1,\cdots,m_L}~\exp \left(f_{m_0,m_1,\cdots,m_L}-\beta_{m_0} 
E_{\bil_m}\right)}}~,
\end{array}
\label{eqn20}
\end{equation}
and 
\begin{equation}
\begin{array}{ll}
\lefteqn{ \exp (-f_{m_0,m_1,\cdots,m_L})}\\
&= \dis{\sum_{E_0,V_1,\cdots,V_L} n(E_0,V_1,\cdots,V_L) \exp \left(-\beta_{m_0} E_{\bil_m}\right)}~. 
\end{array}
\label{eqn21}
\end{equation}
The density of states
$n(E_0,V_1,\cdots,V_L)$ and the dimensionless free energy $f_{m_0,m_1,\cdots,m_L}$ are obtained by solving Eqs.~(\ref{eqn20}) and (\ref{eqn21}) self-consistently by iteration.
The canonical probability distribution at any temperature $T=1/k_{\rm B} \beta$ with any potential-energy parameter value $\bil$  is then given by 
$P(E_0,V_1,\cdots,V_L)=n(E_0,V_1,\cdots,V_L)\exp(-\beta E_{\bil})$. 

Finally, the weight factors for multi-dimensional MUCA (see Eq.~(\ref{eqn4})) 
and multi-dimensional ST (see Eqs.~(\ref{eqn8}) and (\ref{eqn10})) 
are obtained from the generalized density of states $n(E_0,V_1,\cdots,V_L)$ and the dimensionless free energy $f_{m_0,m_1,\cdots,m_L}$, respectively. 
\begin{figure}[b]
\includegraphics[height=3.0cm]{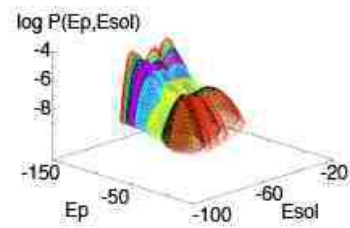}
\caption{Canonical distributions $P(E_{\rm P},E_{\rm SOL})$ with 
the 32 possible parameter sets $(T_{m_0},\lambda_{m_1})$, 
which were obtained by the short two-dimensional REM simulation.}
\label{Fig1}
\end{figure}

As an example of the applications of the present formulations, we now present 
the results of a two-dimensional ST simulation. 
The system is a biomolecluar system studied in Ref. \cite{MSO03b}. 
We set
$E_{\lambda} = E_{\rm P} + \lambda E_{\rm SOL}$, 
where we have $L$ = 1 in Eq. (\ref{eqn1}) and $E_0=E_{\rm P}$ is the conformational energy of the biomolecule and $V_1=E_{\rm SOL}$ is the solvent energy. 
The simulations were started from randomly generated conformations. 
We prepared eight temperatures which are distributed exponentially between $T_1=$ 300 K and $T_8=$ 700 K (i.e., 300.00, 338.60, 382.17, 431.36, 486.85, 549.49, 620.20, and 700.00 K) and four equally-spaced $\lambda$ values ranging from 0 to 1 (i.e., $\lambda_1$ = 0, $\lambda_2$ = 1/3, $\lambda_3$ = 2/3, and $\lambda_4$ = 1). 
The total number of replicas is then 32 ($=8 \times 4$). 

In Fig.~1, the canonical probability distributions at 32 conditions obtained from the two-dimensional REM simulation are shown. For an optimal performance of the REM simulation, there should be enough overlaps between all pairs of neighboring distributions, which will lead to sufficiently uniform and large acceptance ratios of replica exchange. 
We see in Fig.~1 that there are indeed ample overlaps between the neighboring distributions. 
\begin{figure}[t]
(a)
\includegraphics[height=2.4cm]{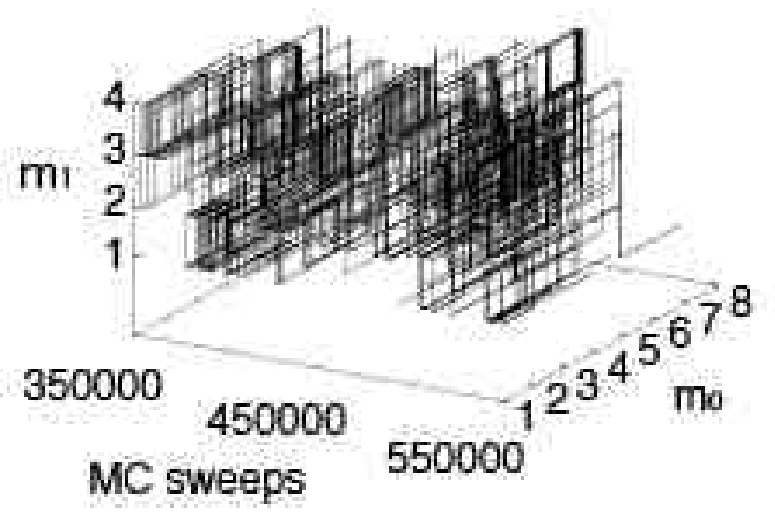}
(b)
\includegraphics[height=2.0cm]{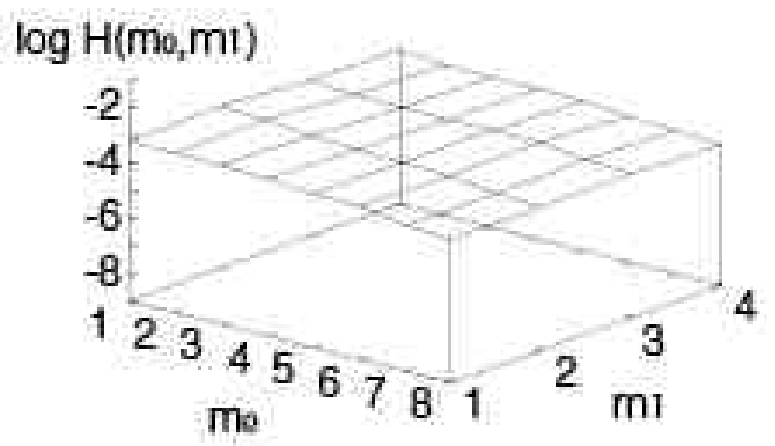}
\caption{Time series (a) and histogram $H(m_0,m_1)$ (b) 
of the parameter labels $m_0$ and $m_1$
for $(T_{m_0},\lambda_{m_1})$, 
which were obtained by the two-dimensional ST simulation.}
\label{Fig3}
\end{figure}

Using the results of this MREM simulation, 
we obtained the two-dimensional ST parameters $f_{m_0,m_1}$ ($m_0=1,\cdots, 8$; $m_1=1,\cdots,4$) by the multiple-histogram reweighting techniques (see Eqs.~(\ref{eqn8}), (\ref{eqn20}), and (\ref{eqn21})), and performed a two-dimensional ST simulation. 

The time series of labels of temperature $T$ and parameter $\lambda$ is shown 
in Fig.~2(a). The random walk in both $T$ space and $\lambda$ space 
was indeed realized. 
The histogram of labels of $T$ and $\lambda$ is shown in Fig.~2(b).
We did get an expected flat histogram in $T$ and $\lambda$. 

Finally, we remark that once the weight factors for the 
multi-dimensional MUCA and ST are obtained, they can give
the weight factors for lower-dimensional cases.
For instance, the weight factor for the {\it multimagnetical algorithm} \cite{MUMA}
can be obtained from that for the
two-dimensional {\it multicanonical-multimagnetical} ensemble
by integrating out the $E_0$ variable (zero-field term).
Likewise, the weight factors for {\it multibaric-multithermal algorithm}
can be reduced to those for {\it multibaric-isothermal} ensemble
and {\it isobaric-multithermal} ensemble \cite{OO}. \\

\noindent
{\bf Acknowledgements}: \\
This work was supported, in part, by Grants-in-Aid
for Scientific Research in Priority Areas (``Water and Biomolecules''
and ``Molecular Theory for Real Systems'')
and for the Next Generation Super Computing Project, Nanoscience Program
from the Ministry of
Education, Culture, Sports, Science and Technology (MEXT), Japan.

\end{document}